\documentclass[aps,prc,twocolumn,amsmath,amssymb,floatfix,nobalancelastpage]{revtex4}
\usepackage{microtype}
\usepackage{newtxtext}
\usepackage[upint,varbb]{newtxmath}
\usepackage{eucal}
\usepackage{graphicx}
\usepackage[caption=false]{subfig}
\usepackage{amsfonts}
\usepackage{amsmath}
\usepackage{amssymb}
\usepackage{textcomp}
\usepackage{color}
\usepackage{bm}
\usepackage{float}
\usepackage[capitalize]{cleveref}
\usepackage{soul}
\usepackage{csquotes}
\usepackage{tikz}
\usepackage{pgfplots}

\pgfkeys{/pgf/number format/.cd,fixed,precision=3}
\pgfplotsset{compat=1.3, every axis/.append style={scale only axis, axis on top,height=(\columnwidth-\columnsep)/2-0.5cm, width=(\columnwidth-\columnsep)/2-0.5cm}}

\definecolor{DarkRed}{rgb}{0.65,0,0}%
\definecolor{Green}{rgb}{0,0.5,0.3}
\definecolor{Purple}{rgb}{0.3,0,0.65}
\definecolor{DarkGray}{rgb}{0.7,0.7,0.7}
\definecolor{Blue}{rgb}{0,0,1}

\newcommand{\cG}[0]{\check{G}}
\newcommand{\cg}[0]{\check{g}}
\newcommand{\rv}[0]{\vec{r}}
\newcommand{\Rv}[0]{\vec{R}}
\newcommand{\kv}[0]{\vec{k}}
\newcommand{\qv}[0]{\vec{q}}
\newcommand{\cT}[0]{\check{T}}
\newcommand{\ct}[0]{\check{t}}
\newcommand{\imp}[0]{\text{imp}}
\newcommand{\cbg}[0]{\check{\bar{g}}}

\renewcommand{\vec}[1]{\bm{#1}}

\begin{document}
\title{The quasiclassical theory for interfaces with spin--orbit coupling}
\author{Morten Amundsen}
\author{Jacob Linder}
\affiliation{Center for Quantum Spintronics, Department of Physics, Norwegian \\ University of Science and Technology, NO-7491 Trondheim, Norway}

\begin{abstract}
\noindent
In recent years, substantial progress has been made regarding the application of quasiclassical theory on superconducting hybrid structures. This theoretical framework is reliant on a proper set of boundary conditions in order to describe multilayered systems. With the advent of the field of superconducting spintronics, systems which combine heavy metal layers, in which there is large spin--orbit coupling, with ferromagnets have received a great deal of attention, due to their potential for generating long range triplet superconductivity. In contrast to interfaces of strongly spin polarized materials, which are well understood, a quasiclassical theory for interfaces in systems where there is significant spin--orbit coupling does not yet exist. After reviewing the quasiclassical theory for interfaces, we here solve this problem by deriving a new set of boundary conditions which take spin--orbit coupling explicitly into account. We then go on to apply these boundary conditions to a superconductor-ferromagnet (SF) bilayer and an SFS Josephson weak link, demonstrating the emergence of long range triplet superconductivity in these systems. 
\end{abstract}

\maketitle

\noindent
\section{Introduction}
The quasiclassical approximation~\cite{serene_pr_1983,belzig_sm_1999,rammer_rmp_1986} is a versatile tool with which complex quantum mechanical problems can be simplified to such an extent that they become numerically solvable. The main assumption of this approximation is that the relevant quantities under study vary on length scales which are much larger than the Fermi wavelength so that the computationally challenging shorter length scale oscillations may be integrated out. This makes the quasiclassical approximation particularly well suited for superconductors, in which the superconducting condensate may remain correlated over mesoscopic distances. Indeed, the most general theoretical framework used to describe superconductors, the Green function technique, requires the solution of the Gor'kov equation~\cite{AGD}, which is too cumbersome in all but a few select problems. Instead, progress can be achieved with its quasiclassical equivalents, the Eilenberger~\cite{eilenberger_zp_1968} and the Usadel equation~\cite{usadel_prl_1970}, which govern quassiclassical Green functions describing only the envelopes of the original propagators, and remain the only viable solution method for many problems of practical interest.     

When non-superconducting materials are attached to a superconductor to form superconducting hybrid structures, an interesting phenomenon occurs. In such systems, superconducting correlations may leak into the adjoining non-superconducting materials, so that they too attain superconducting properties. This is known as the proximity effect. The study of such systems necessarily involves the proper treatment of interfaces between materials. However, while the bulk properties of superconductors are easily described within quasiclassical theory, interfaces between materials is another matter entirely. In the vicinity of an interface, the governing Hamiltonian changes abruptly, which invalidates the use of the quasiclassical approximation. The consequence of this is that the quasiclassical Green functions feature a discontinuous jump at interfaces, the size of which is impossible to determine within quasiclassical theory---additional information is needed. This jump was first computed for ballistic SN structures using a full microscopic description of the interface~\cite{zaitsev_jetp_1984}, thereby giving a set of boundary conditions linking the two materials. These boundary conditions were generalized with the use of a projection operator method~\cite{yip_jltp_1997,eschrig_prb_2000,shelankov_prb_2000}. Alternative derivations, where the interface is treated perturbatively via a $T$ matrix approach~\cite{ashauer_jltp_1986,nagai_jltp_1988,graser_prb_2007} have also been proposed. In the diffusive limit, boundary conditions may be arrived at by connecting the momentum independent diffusive Green functions far away from the interface to ballistic Green functions present in a region immediately surrounding the interface~\cite{kupriyanov_jetp_1988,nazarov_sm_1999}.  

For hybrid structures involving strongly polarized magnetic materials, the interfacial boundary conditions generally become spin active. Such boundary conditions have been formulated heuristically using a tunnelling Hamiltonian~\cite{bergeret_prb_2012}. Another, more fundamental, approach is to connect the two sides of the interface by means of scattering or transfer matrices, in which the spin dependence of the scattering processes are taken into account~\cite{millis_prb_1988,tokuyasu_prb_1988,cottet_prb_2009}. Both the projection operator and the $T$ matrix method have also been successfully generalized to handle spin active interfaces in ballistic systems~\cite{cuevas_prb_2001,eschrig_prl_2003,kopu_prb_2004,fogelstrom_prb_2000,zhao_prb_2004}. The latter method was also applied to diffusive systems in Ref.~\cite{eschrig_njp_2015}, in which a completely general theory for boundary conditions in spin polarized hybrid structures was derived.

When hybrid structures are made from superconducting and ferromagnetic materials, the proximity effect allows for the coexistence of both magnetic and superconducting correlations. This produces a number of interesting effects. One of the more fascinating effects is perhaps the appearance of triplet superconductivity, due to the Zeeman splitting endowing the Cooper pairs with a net momentum~\cite{buzdin_rmp_2005}. In homogeneous ferromagnets, the triplet Cooper pairs remain in the spinless state, and hence the magnetization has a strong depairing effect. In ferromagnets where the magnetization is inhomogeneous, on the other hand, the triplet Cooper pairs may be converted to an equal spin state~\cite{bergeret_prl_2001,bergeret_rmp_2005}, giving them a net spin. In such a state, the Cooper pairs are insensitive to a parallel magnetization, and may therefore persist for long distances into proximitized ferromagnets, a phenomenon known as long range triplet superconductivity. In addition, since these correlations are spin polarized, they may carry spin currents, a realisation which has strongly contributed to the field of superconducting spintronics~\cite{linder_np_2015,eschrig_rpp_2015}. 

Another avenue towards triplet superconductivity is by introducing layers with spin--orbit coupling~\cite{gorkov_prl_01, annunziata_prb_12, bergeret_prb_2014}. In such systems, spin polarized supercurrents may be generated even in homogeneous ferromagnets, which is advantageous from an experimental point of view. However, while the boundary conditions for spin polarized systems is well understood, a theory for interfaces in which spin--orbit coupling is prominent does not yet exist. In this paper, we seek to remedy this by deriving a new set of boundary conditions which take spin--orbit coupling into account, using the general framework of Ref.~\cite{eschrig_njp_2015}. This result consequently allows for a proper treatment of spin-orbit coupled interfaces in quasiclassical theory, which is in principle relevant for any heterostructure, since interfaces break inversion symmetry, but particularly so for heterostructures with heavy metal interlayers.

We here state our main analytical result and comment briefly on the qualitative physical meaning of each term in the boundary condition. Below, we shall derive this result rigorously and provide a clear description of how the various terms arise, and the meaning of each of the symbols.
\begin{align}
&\hat{n}\cdot\cg_1\nabla\cg_1 = T\left[\cg_1\;,\;\cg_2\right] + T_{\alpha}\left[\cg_1\;,\;\check{\vec{\sigma}}_{||}\cg_2\check{\vec{\sigma}}_{||}\right] \nonumber \\
&+T_{\alpha}'\left[\cg_1\;,\;\check{\vec{\sigma}}_{||}\cg_1\check{\vec{\sigma}}_{||}\right]+i\sqrt{T_{\alpha}''T}\left[\cg_1,\left\{\cg_2\left[\check{\vec{\sigma}}_{||},\cg_2\right],\check{\vec{\sigma}}_{||}\right\}\right]\nonumber \\
&+i\sqrt{T_{\alpha}''T}\left[\cg_1\,,\,\left\{\cg_2,\check{\vec{\sigma}}_{||}\right\}\cg_1\check{\vec{\sigma}}_{||} + \check{\vec{\sigma}}_{||}\cg_1\left\{\cg_2,\check{\vec{\sigma}}_{||}\right\}\right]. 
\end{align}
The usual Kupriyanov-Lukichev term \cite{kupriyanov_jetp_1988} is the one proportional to $T$. The term proportional to $T_\alpha$ represents a correction to the usual tunneling boundary condition from the spin-orbit coupling part of the tunneling matrix. The term $T_\alpha'$ represents spin-dependent phase-shifts occurring due to spin-orbit coupling at the interface and thus exists even in the absence of any tunnelling. The terms $\sqrt{T_{\alpha}''T}$ are higher order corrections which exist only in the presence of spin-independent tunneling, spin-orbit coupled tunneling, and spin-orbit coupled reflection.

The quasiclassical theory for boundary conditions is quite intricate, involving many details, so to ensure complete clarity of the ensuing derivation, we include in \cref{sec:review} a review of the treatment of interfaces within quasiclassical theory. In \cref{sec:soc} we formulate the boundary conditions in the presence of spin--orbit coupling, and in \cref{sec:app} we apply the new boundary conditions to example problems, in order to demonstrate their use.

\section{Review of the quasiclassical theory for interfaces}
\label{sec:review}
In this section, a review of quasiclassical boundary conditions will be given, starting with a brief excursion into general quasiclassical theory. 
\subsection{Quasiclassical equations of motion}
All physical observables of interest may be expressed in terms of Green functions, and we use the Keldysh formalism, in which the Green functions take the form of $8\times 8$ matrices in spin$\,\otimes\,$Nambu$\,\otimes\,$Keldysh space, defined as
\begin{align}
\cG = \begin{pmatrix} \hat{G}^R & \hat{G}^K \\ 0 & \hat{G}^A\end{pmatrix}
\quad,\quad
\hat{G}^R = \begin{pmatrix} G^R & F^R \\ \left(F^R\right)^* & \left(G^R\right)^* \end{pmatrix},
\end{align}
where $\hat{G}^A = -\hat{\rho}_3\left(\hat{G}^R\right)^{\dagger}\hat{\rho}_3$, with $\hat{\rho}_3 = \text{diag}\left(+1,+1,-1,-1\right)$, and $\hat{G}^K = \hat{G}^R\hat{h} - \hat{h}\hat{G}^A$ for a given distribution function $\hat{h}$ \cite{belzig_sm_1999}. Furthermore, $G^X$ and $F^X$, with $X\in\left\{R, A, K\right\}$ are $2\times 2$ matrices in spin space. We assume time translation invariance, and Fourier transform in the relative time coordinate, so that we may write $\cG = \cG(\rv_0,\rv_n;\varepsilon)\equiv\cG(\rv_0,\rv_n)$, where $\varepsilon$ is the quasiparticle energy. The equation of motion may then be written as
\begin{align}
\left[\varepsilon\hat{\rho}_3 - \frac{1}{2m}\left(-i\nabla_0\check{I} - \check{\vec{A}}(\rv_0)\right)^2 + \mu\check{I} - \check{\Sigma}(\rv_0)\right.  \nonumber\\
\left.-\vphantom{\frac{1}{2m}} V_{\imp}(\rv_0)\check{I}\right]\cG(\rv_0,\rv_n;\varepsilon)=\delta(\rv_0 - \rv_n)\check{I},
\label{eq:gorkov}
\end{align}
where $\check{I}$ is the $8\times 8$ identity matrix, $\check{\vec{A}}(\rv_0)$ is the vector potential, $V_{\imp}(\rv_0)$ is the impurity potential, and $\check{\Sigma}(\rv_0)$ encompasses any other local self energies, such as the superconducting gap, or an exchange field. The impurity potential in the conventional way---as a perturbation in momentum space. Towards that end, \cref{eq:gorkov} is reformulated as
\begin{align}
&\cG(\kv_0,\kv_n) = \cG_0(\kv_0,\kv_n) \nonumber\\
&+ \int \frac{d\kv_1}{(2\pi)^3}\int \frac{d\kv_2}{(2\pi)^3}\;\cG_0(\kv_0,\kv_1)V_{\imp}(\kv_1 - \kv_2)\cG(\kv_2,\kv_n),
\label{eq:kgorkovint}
\end{align}
where the Fourier transform of the Green function is defined as
\begin{align}
\cG(\kv_0,\kv_n) = \int d\rv_0\int d\rv_n \;\cG(\rv_0,\rv_n)e^{-i\kv_0\cdot\rv_0 + i\kv_n\cdot\rv_n}.
\label{eq:FT}
\end{align}
Note that by changing variables to $\Rv = (\rv_0 + \rv_n)/2$ and $\rv = \rv_0 - \rv_n$, the Fourier transform may also be written as
\begin{align}
\cG(\kv_0,\kv_n) = \int d\Rv \;\cG(\kv,\Rv)e^{-i\Delta\kv\cdot\Rv},
\end{align}
with $\kv = (\kv_0 + \kv_n)/2$ and $\Delta\kv = \kv_0 - \kv_n$, and
\begin{align}
\cG(\kv,\Rv) = \int d\rv\;\cG(\rv,\Rv) e^{-i\kv\cdot\Rv}.
\label{eq:mixedrep}
\end{align}
\cref{eq:mixedrep} is known as the \emph{mixed representation} of $\cG$, as it involves both the center of mass position $\Rv$ and the center of mass momentum $\kv$. The unperturbed Green function $\cG_0(\kv_0,\kv_n)$ satisfies
\begin{align*}
\int \frac{d\kv_1}{(2\pi)^3}\;&\left[\left(\varepsilon\rho_3 - \frac{k_0^2}{2m} + \mu\check{I}\right)\delta(\kv_0 - \kv_1) +\frac{1}{m}\kv_0\cdot\check{\vec{A}}(\kv_0 - \kv_1)\right.\nonumber\\
&\left. - \frac{1}{2m}\left[\check{\vec{A}}(\kv_0 -\kv_1)\right]^2 - \check{\Sigma}(\kv_0 - \kv_1)\vphantom{\left(\varepsilon\rho_3 - \frac{k_0^2}{2m} + \mu\check{I}\right)}\right]\cG_0(\kv_1,\kv_n)\\
& = \check{I}\delta(\kv_0 - \kv_n).
\end{align*}
By making use of standard techniques for diagram summation \cite{AGD}, taking into consideration that $\cG_0(\kv_0,\kv_n)$ depends on two momentum indices, \cref{eq:kgorkovint} can be written as
\begin{align}
&\cG(\kv_0,\kv_n) = \cG_0(\kv_0,\kv_n) \nonumber\\
&+ \int \frac{d\kv_1}{(2\pi)^3}\int \frac{d\kv_2}{(2\pi)^3}\;\cG_0(\kv_0,\kv_1)\check{\Sigma}_{\imp}(\kv_1, \kv_2)\cG(\kv_2,\kv_n),
\label{eq:kgorkovint2}
\end{align}
with the impurity self energy $\check{\Sigma}(\kv_0,\kv_n)$ defined as
\begin{align}
&\check{\Sigma}_{\imp}(\kv_0,\kv_n) = V_{\imp}(\kv_0 - \kv_n) + \nonumber\\
&\int \frac{d\vec{k}_1}{(2\pi)^3}\int \frac{d\vec{k}_2}{(2\pi)^3}\;V_{\imp}(\kv_0 - \kv_1)\cG_0(\kv_1,\kv_2)\check{\Sigma}_{\imp}(\kv_2,\kv_n).
\label{eq:selfenergy}
\end{align}
It is customary to employ the self-consistent Born approximation, in which \cref{eq:selfenergy} is truncated at second order in $V_{\imp}$, and the replacement $\cG_0\to\cG$ is made on the right hand side. The latter is equivalent to including a much greater number of diagrams in the self energy. The impurity potential is assumed to consist of a large number of randomly distributed identical impurities, so that $V_{\imp}(\rv) = \sum_j U(\rv - \rv_j)$. After impurity averaging, ignoring terms which only depend on $U(0)$ since these simply renormalize the chemical potential, the self energy becomes
\begin{align}
\check{\Sigma}_{\imp}(\kv_0,\kv_n) = \int \frac{d\qv}{(2\pi)^3}\;|U(\kv_0 - \qv)|^2\cG(\qv,\qv - \Delta\kv).
\end{align}
In the mixed representation, arrived at by Fourier transforming in the relative momentum $\Delta\kv$, this expression becomes particularly simple,
\begin{align}
\check{\Sigma}_{\imp}(\kv,\Rv) = \int \frac{d\qv}{(2\pi)^3}\;|U(\kv - \qv)|^2\cG(\qv,\Rv).
\end{align}
By comparing with the full Fourier transform, given in \cref{eq:FT}, we may identify $\Rv$ as the center of mass position. \cref{eq:gorkov} takes the following form in the mixed representation,
\begin{align}
\left[\varepsilon\hat{\rho}_3 - \frac{1}{2m}\left(-i\left(i\kv + \frac{1}{2}\nabla_R\right)\check{I} - \check{\vec{A}}(\Rv)\right)^2 + \mu\check{I} - \check{\Sigma}(\Rv)\right.  \nonumber\\
\left.-\vphantom{\frac{1}{2m}} \check{\Sigma}_{\imp}(\kv,\Rv)\right]\otimes\cG(\kv,\Rv)=\check{I},
\label{eq:mrgorkov}
\end{align}
where the operator $\otimes$ indicates the Moyal product, which is defined as $A(\kv,\Rv)\otimes B(\kv,\Rv) = e^{i(\nabla^{(A)}_R\cdot\nabla^{(B)}_k - \nabla^{(A)}_k\cdot\nabla^{(B)}_R)/2}A(\kv,\Rv)B(\kv,\Rv)$. If the spatial variation of $A(\kv,\Rv)$ and $B(\kv,\Rv)$ is slow, one may approximate $A(\kv,\Rv)\otimes B(\kv,\Rv) \simeq A(\kv,\Rv)B(\kv,\Rv)$.

\cref{eq:mrgorkov} may be further simplified by introducing the quasiclassical approximation, wherein the rapid oscillations of the Green function are integrated out~\cite{rammer_rmp_1986},
\begin{align}
\cg(\kv_F,\vec{R}) = \frac{i}{\pi}\int d\xi_k\;\cG(\kv,\vec{R}),
\label{eq:qc}
\end{align}
where $\xi_k = \frac{1}{2m}(\kv^2 - k_F^2)$. In \cref{eq:qc} there is an implicit assumption that the Green function $\cG(\kv,\Rv)$ is strongly peaked at the Fermi level $k_F$, so that only the angular dependence of the momentum $\kv$ appears in the quasiclassical Green function $\cg(\kv_F,\Rv)$. This is satisfied as long as the spatial variation of the self-energies appearing in $\cG$ is sufficiently slow. The quasiclassical approximation may not be applied to \cref{eq:mrgorkov} directly, as it contains both constant terms and terms proportional to $\xi_k$. These terms can be removed by employing the so-called \enquote{left--right} trick~\cite{eilenberger_zp_1968,larkin_jetp_1969,serene_pr_1983}, where one instead considers the difference between \cref{eq:mrgorkov} and its adjoint, thereby cancelling out the problematic terms. Doing so leads to the Eilenberger equation~\cite{eilenberger_zp_1968},
\begin{align}
&i\vec{v}_F\cdot\tilde{\nabla}\cg(\kv_F,\rv) \nonumber \\
&+ \left[\varepsilon\check{\rho}_3 + \check{\Sigma}(\kv_F,\rv) - \check{\Sigma}_{\imp}(\kv_F,\rv)\;,\;\cg(\kv_F,\rv)\right] = 0.
\label{eq:eilenberger}
\end{align}
\cref{eq:eilenberger} is accompanied by a normalization condition on the quasiclassical Green function, $\cg^2 = \check{I}$. 

In the limit of large concentrations of impurities, the effect of frequent scatterings may be included by averaging over momentum direction. This defines a diffusive Green function $\cg_d = \langle\cg\rangle$, and its governing equation of motion, the Usadel equation~\cite{belzig_sm_1999,rammer_rmp_1986,usadel_prl_1970}
\begin{align}
D\nabla\cdot\cg_d\nabla\cg_d + i\left[\varepsilon\rho_3 - \check{\Sigma}\;,\;\cg_d\right] = 0,
\label{eq:usadel}
\end{align}
where $D$ is the diffusion constant.

\subsection{Distinguished impurities}
We next consider a case where there is an additional impurity $\check{V}(\rv_0)$ present, which may in some way be distinguished from the averaged impurities described by $\check{\Sigma}_{\imp}$. We further assume that this impurity is strongly localized at some position. This means that impurity averaging is not possible. Even so, the quasiclassical formulation of the equation of motion for such a system may be arrived at by perturbation theory~\cite{thuneberg_prb_1984}.
Indeed, if any interference between the averaged and the localized impurity is neglected, the integral equation for the Green function once again takes the form of \cref{eq:kgorkovint}, where $\check{V}$ replaces $V_{\imp}$ as the perturbing potential, and $\cG_0$ is the Green function for a system where $\check{\Sigma}_{\imp}$ is included, but where $\check{V} = 0$. By repeated iteration of this equation, it is seen that it may be written in the form
\begin{align}
&\cG(\kv_0,\kv_n) = \cG_0(\kv_0,\kv_n) \nonumber\\
&+ \int \frac{d\kv_1}{(2\pi)^3}\int \frac{d\kv_2}{(2\pi)^3}\;\cG_0(\kv_0,\kv_1)\cT(\kv_1,\kv_2)\cG_0(\kv_2,\kv_n),
\label{eq:ktmat}
\end{align}
with the $T$ matrix defined as
\begin{align}
&\cT(\kv_0,\kv_n) = \check{V}(\kv_0 - \kv_n)\nonumber \\  &+\int\frac{d\kv_1}{(2\pi)^3}\int\frac{d\kv_2}{(2\pi)^3}\;\check{V}(\kv_0-\kv_1)\cG_0(\kv_1,\kv_2)\cT(\kv_2,\kv_n).
\label{eq:tmatrix}
\end{align}

If the distinguished impurity is localized at a position $\Rv_0$, the Fourier transformed of the impurity potential is given as $\check{V}(\qv) = \check{V}_0(\qv) e^{-i\qv\cdot\Rv_0}$, where $\check{V}_0(\qv)$ is a slowly varying function of $\qv$. By inserting this into \cref{eq:tmatrix}, it is seen that the $T$ matrix can be written as
\begin{align}
\cT(\kv_0,\kv_n) = \cT_0(\kv_0,\kv_n) e^{-i(\kv_0 - \kv_n)\cdot\Rv_0},
\label{eq:tt0}
\end{align}
where $\cT_0(\kv_0,\kv_n)$ has the exact same form as \cref{eq:tmatrix}, with the replacements $\check{V}\to\check{V}_0$ and $\check{T}\to\check{T}_0$. It is thus a slowly varying function of $\kv_0$ and $\kv_n$, so that we may approximate $\cT_0(\kv_0,\kv_n)\simeq\cT_0(\kv,\kv)$, where $\kv$ is the center of mass momentum. In the mixed representation, the $T$ matrix then becomes
\begin{align}
\cT(\kv,\Rv) = \cT_0(\kv,\kv)\delta(\Rv - \Rv_0).
\end{align}
The equation of motion in the mixed representation becomes identical to \cref{eq:mrgorkov}, but with the addition of a term $\cT(\kv,\Rv)\otimes\cG_0(\kv,\Rv)$, which is once again approximated by a product. Following the same steps used in deriving \cref{eq:eilenberger} then gives
\begin{align}
&i\vec{v}_F\cdot\tilde{\nabla}\cg(\kv_F,\Rv)+ \left[\varepsilon\check{\rho}_3 + \check{\Sigma}(\kv_F,\Rv) - \check{\Sigma}_{\imp}(\kv_F,\Rv)\;,\;\cg(\kv_F,\Rv)\right] \nonumber \\ &=\left[\ct_0(\kv_F,\Rv)\;,\;\cg(\kv_F,\Rv)\right]\delta(\Rv-\Rv_0),
\label{eq:teilenberger}
\end{align}
where $\ct_0(\kv_F, \Rv)$ is the quasiclassical version of the $T$ matrix, given as
\begin{align}
\ct_0(\kv_F,\Rv) = \check{V}_0(0) + N_0\int\frac{d\Omega_q}{4\pi}\;\check{V}_0(\kv_F - \qv_F)\,\cg(\qv_F,\rv)\,\ct_0(\qv_F,\Rv).
\end{align}

\subsection{Interface}
An interface, which is a plane in three dimensions, may be treated as an impurity that is localized along a specific direction, having translation invariance along the two orthogonal directions. This implies that a similar perturbation expansion as was discussed in the previous section may be applied also in this case. However, the interface may not be constructed from an ensemble of point impurities satisfying \cref{eq:teilenberger} \cite{thuneberg_prb_1984}. This is because i) the pointlike nature of the impurity was explicitly made use of in \cref{eq:tt0}, and ii) interference between different impurities is neglected. Instead, we follow Ref.~\cite{buchholtz_zp_1979} and consider a model surface of the form $\check{V}(\rv_0) = \check{V}_0\delta[\hat{n}\cdot(\rv_0 - \Rv_n)]$, where $\hat{n}$ is the normal vector of the surface, and $\Rv_n$ is a point on the surface. To simplify the notation, we define $\hat{n}\cdot\rv_0 = r_{\perp}$, and $\hat{n}\cdot\Rv_n = R_0$. Furthermore, we have that $\rv_0 = r_{\perp}\hat{n} + \rv_{||}$. Insertion into \cref{eq:tmatrix} allows us to define
\begin{align}
\cT(\kv_0,\kv_n) = \cT_s(\kv_{0,||},\kv_{n,||})e^{-i(k_{0,\perp} - k_{n,\perp})R_0},
\label{eq:TQ}
\end{align}
with $k_{j,\perp}$ and $\kv_{j,||}$, respectively, the orthogonal and parallel components of momentum $j$, with respect to the surface. Moreover,
\begin{align}
\cT_s(\kv_{0,||},\kv_{n,||}) &= \check{V}_0\,(2\pi)^2\delta(\kv_{0,||}-\kv_{n,||}) \nonumber \\
&+ \check{V}_0\int \frac{d\qv_{||}}{(2\pi)^2}\;\check{Q}(\kv_{0,||},\qv_{||})\cT_s(\qv_{||},\kv_{n,||}),
\end{align}
where
\begin{align}
&\check{Q}(\kv_{0,||},\qv_{||})= \nonumber \\
&\int \frac{dk_{1,\perp}}{2\pi}\int \frac{dk_{2,\perp}}{2\pi} \cG_0(\kv_{0,||}+k_{1,\perp}\hat{n},\qv_{||} + k_{2,\perp}\hat{n})e^{i(k_{1,\perp} - k_{2,\perp})R_0} \nonumber\\
&=\int \frac{dq_{\perp}}{2\pi}\;\cG_0(\kv_{0,||},\qv_{||};\qv_{\perp},R_0).
\end{align}
In the mixed represantion, \cref{eq:TQ} simply becomes
\begin{align}
\cT(\kv,\Rv) = \cT_s(\kv_{||},\Rv_{||})\,\delta(R_{\perp}-R_0),
\end{align}
with
\begin{align}
\cT_s(\kv_{||},\Rv_{||}) =& \check{V}_0 + \check{V}_0\,\check{Q}(\kv_{||},\Rv_{||})\otimes\cT_s(\kv_{||},\Rv_{||}) \nonumber \\
\simeq&\check{V}_0 + \check{V}_0\,\check{Q}(\kv_{||},\Rv_{||})\,\cT_s(\kv_{||},\Rv_{||}),
\end{align}
and
\begin{align}
\check{Q}(\kv_{||},\Rv_{||}) = \int \frac{dq_{\perp}}{2\pi}\;\cG_0(q_{\perp}\hat{n} + \kv_{||},R_0\hat{n}+\Rv_{||})
\label{eq:Q}
\end{align}
To find the quasiclassical version of \cref{eq:Q}, we insert the inverse of \cref{eq:qc}, namely $\cG(\kv,\Rv) = -i\pi\cg(\kv_F,\Rv)\,\delta\left(\frac{1}{2m}(\kv^2 - k_F^2)\right)$. Performing the integral over $q_{\perp}$ gives
\begin{align}
\check{Q}(\kv_{||},\Rv_{||}) = -\frac{i}{|v_n|}\cbg(\kv_{||},\Rv_{||}),
\end{align}
with $v_n = \frac{\kv_F\cdot\hat{n}}{m}$,
\begin{align}
\cbg(\kv_{||},\Rv_{||}) = \frac{1}{2}\left[\cg(\kv_{+},\Rv_{||} + R_0\hat{n}) + \cg(\kv_{-},\Rv_{||} + R_0\hat{n})\right],
\end{align}
and $\kv_{\pm} = \pm\sqrt{k_F^2 - \kv_{||}^2}\hat{n} + \kv_{||}$. This means that the quasiclassical $T$ matrix for an interface,
\begin{align}
\ct_s(\kv_{||},\Rv_{||}) = \check{V}_0 - \frac{i}{|v_n|}\check{V}_0\,\cbg(\kv_{||},\Rv_{||})\,\ct_s(\kv_{||},\Rv_{||})
\label{eq:tmatrixsurf}
\end{align}
only depends on the \emph{average} of Green functions whose normal component of the momentum direction point, respectively toward and away from the interface. The equation of motion, takes the same form as \cref{eq:teilenberger}, with the replacements $\ct_0\to\ct_s$ and $\delta(\Rv - \Rv_0)\to\delta(R_{\perp} - R_0)$.

\subsection{Formulation of boundary conditions}
We next want to consider an interface between two different materials. This is done by expanding Hilbert space into two domains, which represent the two sides of the interface. The Green function in this space can be written as
\begin{align}
\breve{g} = \begin{pmatrix} \cg_{11} & \cg_{12} \\ \cg_{21} & \cg_{22} \end{pmatrix},
\label{eq:Gbig}
\end{align}
where the subscripts ``1'' and ``2'' indicate the two materials. The matrices $\cg_{12}$ and $\cg_{21}$ contain creation and annihilation operators on both sides of the interface. These quantities are drone amplitudes that do not have physical meaning, and since they will be eliminated from the theory, we do not specify them further. The interface itself is described as an infinite surface located at some position, and which mediates tunnelling between its two sides. Such a potential may be described as

\begin{align}
\breve{V}_0 = \begin{pmatrix} 0 & \check{V}_{0} \\ \check{V}_{0} & 0\end{pmatrix}.
\end{align}
The surface is treated as a perturbation, and the $T$ matrix is given by \cref{eq:tmatrixsurf}. 

Without the presence of the interface potential, there is no coupling between the two sides. The unperturbed Green function, $\breve{g}_0$, therefore takes the form
\begin{align}
\breve{g}_0 = \begin{pmatrix}\cg_{0,1} & 0 \\ 0 & \cg_{0,2}\end{pmatrix}.
\label{eq:bg0}
\end{align}
Note that \cref{eq:bg0} satisfies a generalized version of \cref{eq:eilenberger}, given as
\begin{align}
i\kv_F\cdot\bar{\nabla}\breve{g}_0 + \left[\breve{\Xi}\;,\;\breve{g}_0\right] = 0,
\label{eq:be}
\end{align}
where
\begin{align}
\breve{\Xi} = \begin{pmatrix}\varepsilon\check{\rho}_3 + \check{\Sigma}_1 - \check{\Sigma}_{\imp} & 0 \\ 0 & \varepsilon\check{\rho}_3 + \check{\Sigma}_2 - \check{\Sigma}_{\imp}\end{pmatrix}.
\end{align}
Similarly, $\breve{g}$ satisfies
\begin{align}
i\kv_F\cdot\bar{\nabla}\breve{g} + \left[\breve{\Xi}\;,\;\breve{g}\right] = \left[\breve{t}_s\;,\;\breve{g}_0\right]\delta(R_{\perp} - R_0).
\label{eq:bet}
\end{align}

To find a relationship between $\breve{g}$ and $\breve{g}_0$, \cref{eq:bet} may be integrated along a small interval surrounding $R_0$. For a trajectory (as determined by $\kv_F$) crossing the interface, this leads to
\begin{align}
\breve{g}(R_0^+) - \breve{g}(R_0^-) = \frac{1}{i\kv_F\cdot\hat{n}}\left[\breve{t}_s\;,\;\breve{g}_0(R_0)\right],
\label{eq:gbjump}
\end{align}
where we henceforth define $\hat{n}$ to be an outwards pointing surface normal. While \cref{eq:Gbig} is defined everywhere in space, its diagonal elements, $\check{g}_{11}$ and $\check{g}_{22}$, only make sense physically in, respectively, in material 1 and 2, i.e., on opposite sides of the interface. Without loss of generality, we choose material 1 to be the active material, that is, the material for which we formulate the boundary conditions. This means that $\hat{n}$ points from material~1 to material~2. If $\breve{g}(R_0^-)$ is located in material 1, we thus need to eliminate $\breve{g}(R_0^+)$, which is located in material 2. This can be done by making use of a generalized normalization condition, given by
\begin{align}
\left(\breve{g} + \text{sgn}(\kv_F\cdot\hat{n})\right)\left(\breve{g}_{0} - \text{sgn}(\kv_F\cdot\hat{n})\right) =& 0 \label{eq:gg01}\\
\left(\breve{g}_{0} + \text{sgn}(\kv_F\cdot\hat{n})\right)\left(\breve{g} - \text{sgn}(\kv_F\cdot\hat{n})\right) =& 0,
\label{eq:gg02}
\end{align} 
These conditions are clearly satisfied in the special case of $\breve{g} = \breve{g}_{0}$. That they are valid also in the more general case, can be seen by considering the original derivation of Shelankov~\cite{shelankov_jltp_1985}, introducing the interface and taking care to only evaluate points away from the interface, so that $\breve{g}$ and $\breve{g}_{0}$ both satisfy \cref{eq:be}---albeit with different boundary conditions~\cite{buchholtz_jltp_2002,rainer_book_1988}. In material~1 the Green function which describes a particle on a trajectory towards the interface satisfies $\text{sgn}(\kv_F\cdot\hat{n}) = +1$. We label these Green functions as $\cg_{11}^i$ and $\cg_{0,1}^i$, indicating that they are incoming with respect to the interface. Similarly, Green functions where $\text{sgn}(\kv_F\cdot\hat{n}) = -1$ are labelled as outgoing; $\cg_{11}^o$ and $\cg_{0,1}^o$. By evaluating \cref{eq:gg01,eq:gg02} immediately adjacent to, and on opposite sides of the interface, and inserting \cref{eq:gbjump}, the following boundary conditions may be derived at the interface,
\begin{align}
\cg^i_{11} =& \cg^i_{0,1} + \frac{1}{2i|v_n|}\left(\cg_{0,1}^{i} - \check{I}\right)\check{t}_{s,11}\left(\cg_{0,1}^{i} + \check{I}\right), \label{eq:bcit1}\\
\cg^o_{11} =& \cg^o_{0,1} + \frac{1}{2i|v_n|}\left(\cg_{0,1}^{o} + \check{I}\right)\check{t}_{s,11}\left(\cg_{0,1}^{o} - \check{I}\right).
\label{eq:bcit2}
\end{align} 

Due to the form of the interface potential $\breve{V}_0$, $\breve{t}_s$ is in general dense in 1--2-space. However, a closed solution for the $\check{t}_{s,11}$ element may be found by iterating \cref{eq:tmatrixsurf} once~\cite{cuevas_prb_2001}, 
\begin{align}
\check{t}_{s,11} = -\frac{i}{|v_n|}\check{V}_0\,\cbg_{0,2}\check{V}_0 - \frac{1}{v_n^2}\check{V}_0\,\cbg_{0,2}\check{V}_0\,\cbg_{0,1}\ct_{s,11}.
\label{eq:trec}
\end{align} 
Note that $\cbg_{0,j} = \frac{1}{2}\left(\cg_{0,j}^i + \cg_{0,j}^o\right)$, for a given side of the interface $j$. \cref{eq:trec} may easily be solved for $\ct_{s,11}$, giving
\begin{align}
\check{t}_{s,11} = \frac{1}{i|v_n|}\left[\check{I} + \frac{1}{v_n^2}\check{V}_0\,\cbg_{0,2}\check{V}_0\,\cbg_{0,1}\right]^{-1}\,\check{V}_0\,\cbg_{0,2}\check{V}_0.
\label{eq:tinterface}
\end{align}

To summarize the progress so far, we have found boundary conditions for the quantities $\cg_{11}^i$ and $\cg_{11}^o$, given in \cref{eq:bcit1,eq:bcit2}, expressed entirely in terms of $\cg_{0,j}^{i/o}$. While these unperturbed Green functions exist everywhere in space, they are only physically valid solutions on their respective sides of the interface. Furthermore, as discussed in Ref.~\cite{eschrig_njp_2015}, they can be easily modified to describe a system with an impenetrable interface by having them satisfy the condition
\begin{align}
\cg^o_{0,j} = \check{S}\cg^i_{0,j}\check{S}^{\dagger}
\label{eq:gio}
\end{align}  
where $\check{S}$ is a scattering matrix. For spin independent scattering, $\check{S} = \check{I}$. The point is that rather than considering the interface as a perturbation to omnipresent Green functions, we may exploit the fact that we only evaluate $\cg_{0,j}$ in one region of space to redefine them to represent a system with an impenetrable interface, simply by imposing \cref{eq:gio}. This provides significant benefits. Ultimately, the goal is to derive boundary conditions for diffusive systems, which are governed by diffusive Green functions $\cg_{d,j}$. Even for such a system, there will always be a ballistic zone immediately surrounding an interface, since a particle travelling away from the interface will traverse a distance on the order of the mean free path before encountering its first impurity. Diffusive systems with an interface are therefore governed by ballistic Green functions $\cg_{0,j}^{i/o}$, with $\check{\Sigma}_{\imp} = 0$, close to the interface, and by $\cg_{d,j}$ far away from the interface. In between is an asymptotic matching region known as the \emph{isotropization zone}, within which the momentum dependence of the ballistic Green functions are averaged out by repeated impurity scatterings. Matching the Green functions of the two regimes is only possible if the following conditions are satisfied~\cite{millis_prb_1988, nazarov_sm_1999, cottet_prb_2009},
\begin{align}
\left(\cg^i_{0,j} + \check{I}\right)\left(\cg_{d,j} - \check{I}\right) =& 0, \label{eq:iso1}\\
\left(\cg_{d,j} + \check{I}\right)\left(\cg^i_{0,j} - \check{I}\right) =& 0,\label{eq:iso2} \\
\left(\cg^o_{0,j} - \check{I}\right)\left(\cg_{d,j} + \check{I}\right) =& 0,\label{eq:iso3} \\
\left(\cg_{d,j} - \check{I}\right)\left(\cg^o_{0,j} + \check{I}\right) =& 0.\label{eq:iso4}
\end{align}  
Using \cref{eq:iso1,eq:iso2,eq:iso3,eq:iso4} and \cref{eq:gio}, the ballistic Green functions at the interface may be expressed in terms of their diffusive counterparts as
\begin{align}
\cg^i_{0,j} = &\left(\check{S}^{\dagger}\cg_{d,j}\check{S} + \cg\right)^{-1}\left(2\check{I} + \cg - \check{S}^{\dagger}\cg_{d,j}\check{S}\right),\label{eq:bcd1} \\
\cg^o_{0,j} = &\left(2\check{I} + \cg - \check{S}\cg_{d,j}\check{S}^{\dagger}\right)\left(\check{S}\cg_{d,j}\check{S}^{\dagger} + \cg\right)^{-1}.\label{eq:bcd2}
\end{align} 
The quasiclassical boundary conditions are completed by computing the \emph{matrix current} directed at the interface. In the ballistic zone it is defined as 

\begin{align}
\check{J}_n = \int \frac{d\Omega}{2\pi} \hat{n}\cdot\vec{v}_F\left(\cg^i_{jj} - \cg^o_{jj}\right),
\label{eq:mjb}
\end{align} 
where the integration measure is an angular average over a hemisphere. In the diffusive zone the matrix current is given as~\cite{nazarov_sm_1999}
\begin{align}
\check{J}_n = \sigma_j\hat{n}\cdot\cg_{d,j}\nabla\cg_{d,j},
\label{eq:mjd}
\end{align}
where $\sigma_j$ is the normal state conductivity of material $j$. The matrix current is conserved across the isotropization zone~\cite{cottet_prb_2009}, and hence \cref{eq:mjd} may be equated with \cref{eq:mjb}, which in turn is determined from \cref{eq:bcit1,eq:bcit2} and \cref{eq:bcd1,eq:bcd2}, thus giving the complete boundary conditions solely in terms of the diffusive Green functions.

\begin{figure}[h]
\begin{tikzpicture}
\begin{axis}[samples=100,name=plot1,xmin=0,xmax=pi/2,ymin=0,ymax=1.1,x label style={at={(axis description cs:0.5,-0.1)},anchor=north},
xlabel = {$\theta/\pi$},y label style={at={(axis description cs:-0.05,.5)},anchor=south}, ylabel={$\tau(\theta)/\tau(0)$},ytick={0,1},xtick={0,0.7854,1.5708},xticklabels={0,0.25,0.5}]
\addplot+[mark=none,smooth] {4*cos(deg(x))^2/(1 + cos(deg(x))^2)^2};
\node at (axis cs:pi/2-0.2,1) {\textbf{a)}};
\end{axis}
\begin{axis}[samples=100,name=plot2,at=(plot1.right of south east),anchor=left of south west, xmin=0,xmax=5,ymin=0,ymax=0.2,x label style={at={(axis description cs:0.5,-0.1)},anchor=north},
xlabel = {$V_0/v_F$},y label style={at={(axis description cs:-0.15,.5)},anchor=south}, ylabel={$T$}]
\addplot+[domain=0.001:5,mark=none,smooth] {0.5*x^2*(rad(atan(1/x))/x - 1/(1+x^2))};
\node at (axis cs:4.4,0.18) {\textbf{b)}};
\end{axis}
\end{tikzpicture}
\caption{a) The angular distribution of the coupling constant $\tau(\theta)$, as determined from \cref{eq:tau}. b) The strength of the angularly averaged coupling constant $T$ in the tunnelling limit, as given in \cref{eq:KLT}.}
\label{fig:CC}
\end{figure}
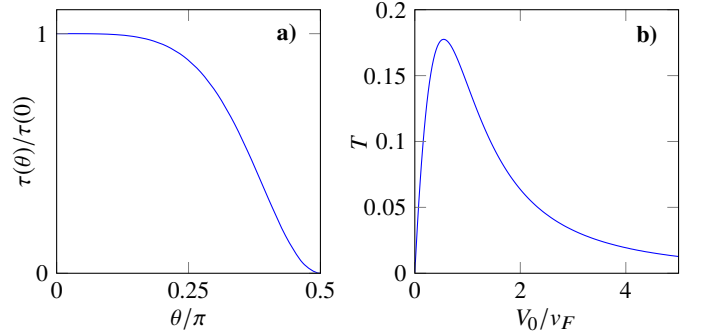

For spin independent scattering we have $\check{S} = \check{I}$, which gives $\cg_{0,j}^i = \cg_{0,j}^o = \cg_{d,j}$. This means that the difference between \cref{eq:bcit1,eq:bcit2} reduces to $\cg^{i}_{11} - \cg^{o}_{11} = \frac{1}{i|v_n|}\left[\ct_{11} , \cg_{d,1}\right]$. By using \cref{eq:tinterface,eq:mjb,eq:mjd} one may, with some algebra~\cite{kopu_prb_2004}, produce Nazarovs boundary conditions~\cite{nazarov_sm_1999},
\begin{align}
\sigma_1\hat{n}\cdot\cg_{d,1}\nabla\cg_{d,1} =\int^{\frac{\pi}{2}}_0d\theta\; \frac{\sin\theta\,\tau(\theta)\left[\cg_{d,1}\;,\;\cg_{d,2}\right]}{4\check{I} + \tau(\theta)\left(\left\{\cg_{d,1}\;,\;\cg_{d,2}\right\} - 2\check{I}\right)},
\label{eq:nazarov}
\end{align}
where the coupling constant $\tau(\theta)$ is given as
\begin{align}
\tau(\theta) = \frac{4\upsilon^2\cos^2\theta}{(\cos^2\theta + \upsilon^2)^2},
\label{eq:tau}
\end{align}
with $\upsilon = V_0/v_F$, and $\theta$ is the angle of incidence with respect to the interface. $\tau(\theta)$ is shown in \cref{fig:CC}a), where it is seen that it attains its maximum value for normal incidence ($\theta = 0$), and goes to zero for trajectories parallel to the interface ($\theta = \frac{\pi}{2}$), which is intuitively reasonable.

When $\tau(\theta)$ is small, e.g., in the tunnelling limit, we may neglect its contribution to the denominator of \cref{eq:nazarov}. This gives the Kupriyanov--Lukichev boundary conditions~\cite{kupriyanov_jetp_1988},
\begin{align}
\hat{n}\cdot\cg_{d,1}\nabla\cg_{d,1} = T\left[\cg_{d,1}\;,\;\cg_{d,2}\right],
\label{eq:KL}
\end{align}
where $T$ is the angular average of the coupling constant, and is given as
\begin{align}
T = \frac{\upsilon^2}{2\sigma_1}\left[\frac{1}{\upsilon}\arctan\left(\frac{1}{\upsilon}\right) - \frac{1}{1+\upsilon^2}\right].
\label{eq:KLT}
\end{align}
$T$ is shown for increasing $V_0$ in \cref{fig:CC}b). For $V_0 = 0$, there is no coupling between the two sides. The system reduces to its unperturbed state in which the interface is impenetrable, and hence $T = 0$. For large values of $V_0$, $T$ also goes to zero. The reason for this is that when the barrier potential increases, incoming particles are more likely to be reflected than transmitted, which means that the two sides eventually become decoupled also in this limit.   

\section{Interfaces with spin--orbit coupling}
\label{sec:soc}
We will now consider an interface to a material within which spin--orbit coupling plays a prominent role, for instance a heavy metal. This means that the transmission probability will depend on both the spin of the incoming particle and its angle of incidence. We model this with a Rashba-like tunnelling coupling, 
\begin{align}\label{eq:vsoc}
\check{V}_0 = w\check{I} + w_\alpha(\hat{n}_{\alpha}\times\hat{\kv}_F)\cdot\check{\vec{\sigma}},
\end{align}
where $\hat{n}_{\alpha}$ is a unit vector indicating the direction in which the symmetry is broken, and $w$ is a spin independent contribution to the tunnelling between the two sides separated by the interface, while $w_\alpha$ is the strength of the spin--orbit coupling contribution to the tunnelling. As the barrier region increases in width, both $w$ and $w_\alpha$ decrease toward zero.

\subsection{Scattering matrix}
As a step towards formulating the boundary conditions, we need to find the scattering matrix for a system where the interface is impenetrable. However, even though the probability for transmission through the interface is zero, an incoming particle may still penetrate the barrier for some distance before being reflected. While inside the scattering region, the particle will experience the spin--orbit coupling and  accumulate a spin dependent phase, the size of which depends on the incoming angle of incidence. To find the scattering matrix for this process, we assume that sufficiently close to the interface, the interface potential is large enough to dominate all other self energies~\cite{millis_prb_1988}. This means that we may follow the procedure of Ref.~\cite{tokuyasu_prb_1988} and describe the interface as a step function potential in a free electron gas. For simplicity, we use a cylindrical coordinate system in which the $z$ axis is aligned with the surface normal. The Hamiltonian for such a system is given as
\begin{align}
H = -\frac{1}{2m}\nabla^2 -\mu+ \left(\mu+\varepsilon_g + i\alpha(\hat{z}\times\nabla)\cdot\vec{\sigma}\right)\theta(z),
\label{eq:SH}
\end{align}
where $\alpha$ expresses the strength of the spin--orbit interaction at the interface, $\mu = \varepsilon_F$ is the chemical potential, and $\theta(z)$ is the Heaviside step function. We consider particles with energies close to the Fermi energy, meaning excitation energies $\varepsilon\simeq 0$, which is the relevant energy regime for quasiclassical theory. The wave function therefore satisfies $H\psi \simeq 0$. The parameter $\varepsilon_g$ expresses an energy gap, and is included to ensure that the wave function is evanescent in the barrier region. Since the interface is assumed to be perfectly smooth, the momentum parallel to the interface, $\kv_{||}$, is conserved during the scattering process. Hence, we may use the ansatz $\psi(\rv) = e^{i\kv_{||}\cdot\rv_{||}}\phi(z)$. \cref{eq:SH} then takes the form
\begin{align}
&\phi''(z) + \left[2m\varepsilon_F -\kv_{||}^2\right]\phi(z) = 0,\quad z < 0 \label{eq:phi1}\\
&\phi''(z) - \left[\kv_{||}^2 + 2m(\varepsilon_g + \alpha(\hat{z}\times\kv_{||})\cdot\vec{\sigma})\right]\phi(z) = 0,\quad z > 0
\label{eq:phi2}
\end{align} 
For $z < 0$, the solution of \cref{eq:phi1} is given as
\begin{align}
\phi(z) = \begin{pmatrix} A_1 \\ A_2\end{pmatrix} e^{ik_{\perp}z} + \begin{pmatrix} B_1 \\ B_2\end{pmatrix} e^{-ik_{\perp}z},
\end{align}
where $k_{\perp} = \sqrt{k_F^2 - \kv_{||}^2}$. For $z > 0$, we get
\begin{align}
\phi(z) = C\begin{pmatrix} 1 \\ -ie^{i\varphi}\end{pmatrix}e^{-q_{+}z} + D\begin{pmatrix}1 \\ ie^{i\varphi}\end{pmatrix}e^{-q_-z},
\end{align}
where $\varphi$ is the azimuthal incidence angle. The momentum $q_{\pm}$ is given as
\begin{align}
q_{\pm} = \sqrt{\kv_{||}^2 + 2m\left(\varepsilon_g \pm \alpha|\kv_{||}|\right)}.
\end{align}

The scattering matrix is found by relating the coefficients $A_{1,2}$ to $B_{1,2}$ via the matrix equation $B = SA$. This is done by enforcing continuity of $\phi$ and $\phi'$ at $z = 0$, and leads to
\begin{align}
S = \frac{1}{(k_{\perp} + iq_+)(k_{\perp} + iq_-)}\begin{pmatrix}k_{\perp}^2 + q_+q_- & e^{-i\varphi}k_{\perp}(q_+ - q_-) \\ e^{i\varphi}k_{\perp}(q_- - q_+) & k_{\perp}^2 + q_+q_- \end{pmatrix}.
\label{eq:Smat}
\end{align}
Since the scattering matrix is unitary, $SS^{\dagger} = I$, it may be parametrized as 
\begin{align}
S = e^{i\beta}e^{i\gamma(\hat{e}\cdot\vec{\sigma})} = e^{i\beta}\left(\cos\gamma + i\hat{e}\cdot\vec{\sigma}\sin\gamma\right),
\end{align}
where $\gamma$ is a spin mixing angle and $\hat{e}$ is a unit vector. By inspection, we see that we may define ${\hat{e}\cdot\vec{\sigma} = -\sigma_x\sin\theta\sin\varphi + \sigma_y\sin\theta\cos\varphi}$.
To find an expression for $\gamma$, we define the constants $a = \sqrt{\frac{\varepsilon_g}{\varepsilon_F}}$ and $b = \frac{2m\alpha}{k_F}$. The former expresses the strength of the barrier potential, and the latter the strength of the spin--orbit coupling. Furthermore, we have $|\kv_{||}| = k_F\sin\theta$, where $\theta \in [0,\pi/2]$ is the polar angle of incidence. We assume that the barrier potential is strong, so that an incoming particle only penetrates a short distance into the scattering region before being reflected, and hence $a \gg b$. From \cref{eq:Smat} we may then identify the spin mixing angle as
\begin{align}\label{eq:spinmixingangle}
\gamma = \arctan\left(\frac{k_{\perp}(q_+ - q_-)}{\sin\theta \left(k_{\perp}^2 + q_+q_-\right)}\right)\simeq K\cos\theta,
\end{align}
with $K = \frac{b}{a^3}$ when we assume that $a\gg 1$, corresponding to a large band gap in the insulator. The parameter $\beta$ describes an overall phase, which is inconsequential for the boundary conditions, and hence we set $\beta = 0$. In coordinate free form, the scattering matrix therefore takes the form,
\begin{align}
S = e^{i\gamma(\hat{n}_{\alpha}\times\hat{k})\cdot\vec{\sigma}}.
\end{align}
As before, $\hat{n}_{\alpha}$ is a unit vector that is either parallel or antiparallel to the interface normal, depending on the direction of symmetry breaking.
In Nambu-space, the scattering matrix becomes
\begin{align}
\hat{S} = \begin{pmatrix} S(\kv) & 0 \\ 0 & S^*(-\kv) \end{pmatrix} = e^{i\gamma(\hat{n}_{\alpha}\times\hat{k})\cdot\hat{\vec{\sigma}}\hat{\rho}_3},
\label{eq:S}
\end{align}
with $\hat{\vec{\sigma}} = \text{diag}\left(\vec{\sigma}\;,\;\vec{\sigma}^*\right)$. Note that the spin mixing angle is antisymmetric in $\kv$, e.g., $\gamma(\kv) = -\gamma(-\kv)$. Finally, the scattering matrix is diagonal in Keldysh space, e.g., $\check{S} = \text{diag}\left(\hat{S}\;,\;\hat{S}\right)$. 
\subsection{Boundary conditions}
Finding the correct boundary conditions has now become a matter of identifying the terms in \cref{eq:mjb}. To achieve this, we include only the lowest order tunnelling contributions, 
\begin{align}
\ct_{s,11}&\simeq \frac{1}{i|v_n|}\check{V}_0\cbg_{0,2}\check{V}_0.\nonumber \\
&=\frac{1}{i|v_n|}\left[w^2\cbg_{0,2} + w_\alpha w\left\{\cbg_{0,2},\check{\zeta}_k\right\} + w_\alpha^2\check{\zeta}_k\cbg_{0,2}\check{\zeta}_k\right], 
\label{eq:tLfirst}
\end{align}
where $\check{\zeta}_k = (\hat{n}_{\alpha}\times\hat{k})\cdot\check{\sigma}\check{\rho}_3$. Furthermore, the spin mixing angle $\gamma$ is assumed to be small, and we therefore keep terms in the scattering matrix only up to second order. This gives

\begin{align}
\check{S} \simeq \left(1-\frac{1}{2}\gamma^2\right)\check{I} + i\gamma\check{\zeta}_k.
\label{eq:Sapprox}
\end{align}
In addition, we approximate 
\begin{align*}
\left[\check{S}^{\dagger}\cg_{d,j}\check{S} + \cg_{d,j}\right]^{-1} &= \frac{1}{2}\left[\check{I} + \frac{1}{2}\left(\cg_{d,j}\check{S}^{\dagger}\cg_{d,j}\check{S} -\check{I}\right) \right]^{-1}\cg_{d,j}, \\
&\simeq \frac{1}{2}\left[\check{I} - \frac{1}{2}\left(\cg_{d,j}\check{S}^{\dagger}\cg_{d,j}\check{S} -\check{I}\right) \right]\cg_{d,j},
\end{align*}
and similarly for $\left[\check{S}\cg_{d,j}\check{S}^{\dagger} + \cg_{d,j}\right]^{-1}$, leading to
\begin{align}
\cg_{0,j}^i \simeq& \frac{1}{2}\check{I} + \frac{3}{2}\cg_{d,j}-\frac{1}{2}\cg\check{S}^{\dagger}\cg_{d,j}\check{S} - \frac{1}{2}\cg_{d,j}\check{S}^{\dagger}\cg_{d,j}\check{S}\cg_{d,j}, \\
 \cg_{0,j}^o \simeq& \frac{1}{2}\check{I} + \frac{3}{2}\cg_{d,j}-\frac{1}{2}\check{S}\cg_{d,j}\check{S}^{\dagger}\cg - \frac{1}{2}\cg_{d,j}\check{S}\cg_{d,j}\check{S}^{\dagger}\cg_{d,j}.
\end{align}
By performing the first order expansion in this way, we ensure that in the limit of spin independent scattering, $\check{S} = \check{I}$, we get $\cg_{0,j}^i = \cg_{0,j}^o = \cg_{d,j}$. By inserting \cref{eq:Sapprox} we get
\begin{align}
\cg^i_{0,j}\simeq& \cg_{d,j} -\frac{1}{2}\gamma^2\left(\cg-\check{I}\right)+ \frac{1}{2}i\gamma\left(\cg_{d,j} - \check{I}\right)\left[\check{\zeta}_k\;,\;\cg_{d,j}\right] \nonumber\\
&-\frac{1}{2}\gamma^2\cg\check{\zeta}_k\cg\check{\zeta}_k\left(\cg + \check{I}\right), \\
\cg^o_{0,j}\simeq& \cg_{d,j} -\frac{1}{2}\gamma^2\left(\cg-\check{I}\right)+ \frac{1}{2}i\gamma\left(\cg_{d,j} + \check{I}\right)\left[\check{\zeta}_k\;,\;\cg_{d,j}\right] \nonumber\\
&-\frac{1}{2}\gamma^2\left(\cg + \check{I}\right)\check{\zeta}_k\cg\check{\zeta}_k\cg.
\end{align}

Finally, we compute the full Green functions from \cref{eq:bcit1,eq:bcit2}, and find the boundary conditions from \cref{eq:mjb,eq:mjd}. Note that due to the angular averaging, all odd terms in $\check{\zeta}_k = (\hat{n}_{\alpha}\times\hat{k})\cdot\check{\sigma}\check{\rho}_3$ cancel, and thus we remove them immediately. Furthermore, the spin--orbit coupling is assumed to stem from the interface, which means that $\hat{n}_{\alpha}$ is parallel to $\hat{n}$. Since only even orders of the former appears, we may set $\hat{n}_{\alpha} = \hat{n}$. The matrix current is then given as
\begin{align}
\sigma_1\hat{n}\cdot\cg_1\nabla\cg_1 = \int \frac{d\Omega}{2\pi}\;\frac{1}{|v_n|}\left[\cg_1\;,\;\check{\mathcal{I}}\right],
\label{eq:Mcurrent}
\end{align} 
with $d\Omega = \sin\theta\;d\theta\;d\phi$, $v_n = v_F\cos\theta$, and

\begin{align}
\check{\mathcal{I}} =& w^2\cg_2 + \check{\zeta}_k\left(w_\alpha^2\cg_2 +\frac{1}{2}\gamma^2\cg_1\right)\check{\zeta}_k \nonumber \\
&+ \frac{1}{2}i\gamma w_\alpha w\left\{\cg_2\left[\check{\zeta}_k,\cg_2\right],\check{\zeta}_k\right\} \nonumber \\
&+\frac{1}{2}i\gamma w_\alpha w\left(\left\{\cg_2,\check{\zeta}_k\right\}\cg_1\check{\zeta}_k + \check{\zeta}_k\cg_1\left\{\cg_2,\check{\zeta}_k\right\}\right),
\end{align}
where we have neglected terms of order $w^2\gamma^2$. From the $\phi$ integration we find that, for an arbitrary matrix $\check{M}$,
\begin{align*}
\check{\zeta}_k\check{M}\check{\zeta}_k = \sin^2\theta\,\check{\vec{\sigma}}_{||}\check{M}\check{\vec{\sigma}}_{||},
\end{align*}
where $\check{\vec{\sigma}}_{||} = [\check{\vec{\sigma}} - \hat{n}\left(\hat{n}\cdot\check{\vec{\sigma}}\right)]\check{\rho}_3$, i.e., only spin directions parallel to the interface contribute to the boundary conditions. 

The $\theta$ integration of the spin independent term in \cref{eq:Mcurrent} diverges.  However, when $\alpha = 0$, we know that including all orders of the $T$ matrix, given in \cref{eq:tinterface}, yields a finite expression---namely, \cref{eq:nazarov}. This means that the divergence appears when the $T$ matrix is truncated to give \cref{eq:tLfirst}. The interpretation of this is that microscopic analytical expressions for the coupling constants due to $w$, $w_\alpha$, and $\gamma$ cannot be found within the present theory, and they instead become input parameters. After the $\theta$ integration we therefore get
\begin{align}
&\hat{n}\cdot\cg_1\nabla\cg_1 = T\left[\cg_1\;,\;\cg_2\right] + T_{\alpha}\left[\cg_1\;,\;\check{\vec{\sigma}}_{||}\cg_2\check{\vec{\sigma}}_{||}\right] \nonumber \\
&+T_{\alpha}'\left[\cg_1\;,\;\check{\vec{\sigma}}_{||}\cg_1\check{\vec{\sigma}}_{||}\right]+i\sqrt{T_{\alpha}''T}\left[\cg_1,\left\{\cg_2\left[\check{\vec{\sigma}}_{||},\cg_2\right],\check{\vec{\sigma}}_{||}\right\}\right]\nonumber \\
&+i\sqrt{T_{\alpha}''T}\left[\cg_1\,,\,\left\{\cg_2,\check{\vec{\sigma}}_{||}\right\}\cg_1\check{\vec{\sigma}}_{||} + \check{\vec{\sigma}}_{||}\cg_1\left\{\cg_2,\check{\vec{\sigma}}_{||}\right\}\right], 
\label{eq:fullbc}
\end{align}
where $T_{\alpha}'' = \frac{1}{2}T_{\alpha}T_{\alpha}'$. The parameter $T$ may be identified by comparing with \cref{eq:KL}, and is hence given by \cref{eq:KLT}. The parameter $T_\alpha$ arises from the spin-orbit coupling part ($w_\alpha)$ of the tunnelling potential in Eq. (\ref{eq:vsoc}) whereas $T_\alpha'$ arises from the interfacial spin-orbit coupling ($\alpha$) giving rise to a spin-mixing angle in Eq. (\ref{eq:spinmixingangle}). \cref{eq:fullbc} is the main result of this paper. 

A special case is worth commenting on. In the absence of any tunnelling, as is the case for a superconductor interfaced by a spin-orbit coupled insulator, only $T_\alpha'$ is non-zero, whereas all other terms vanish, giving the boundary condition
\begin{align}
\hat{n}\cdot\cg_1\nabla\cg_1 =T_{\alpha}'\left[\cg_1\;,\;\check{\vec{\sigma}}_{||}\cg_1\check{\vec{\sigma}}_{||}\right].
\end{align}
This boundary condition could thus be used to look for possible bound states induced at the interface of a superconductor and a spin--orbit coupled insulator. The fact that the $T_\alpha'$ term exists despite the absence of a material to tunnel into on the other side of the interface is clear from the fact that this term only depends on $\check{g}_1$. In this sense, it may be thought of as the spin-orbit coupled equivalent of the spin-dependent phase-shift term $G_\phi$ previously discussed in the context of ferromagnetic insulators \cite{cottet_prb_2009, eschrig_njp_2015}.

\section{Applications}
\label{sec:app}
In the following we will apply the boundary conditions derived in \cref{eq:fullbc} to a set of example problems. In particular, we will consider superconducting hybrid structures in which a non-superconducting material is proximitized to a superconductor. The boundary conditions are assumed to represent a thin intermediary layer of a material with strong spin--orbit coupling. We neglect the inverse proximity effect, in which case the superconductors are approximately described by the Bardeen-Cooper-Schrieffer (BCS) bulk Green function,
\begin{align}
\cg_{\text{BCS}} = \begin{pmatrix} \cosh\theta I & e^{i\phi}\sinh\theta i\sigma_y \\ e^{-i\phi}\sinh\theta i\sigma_y & \cosh\theta I\end{pmatrix},
\label{eq:gbcs}
\end{align}
where $\theta = \arctan\frac{\Delta}{\varepsilon}$ for a given quasiparticle energy $\varepsilon$ and gap size $\Delta$, and $\phi$ is the superconducting phase. In other words, we solve the Usadel equation only in the non-superconducting material. We note that $\left[\cg_{\text{BCS}},\check{\vec{\sigma}}_{||}\right] = 0$, and hence the boundary conditions simplify to
\begin{align}
&\hat{n}\cdot\cg_1\nabla\cg_1 = (T+T_{\alpha})\left[\cg_1\;,\;\cg_2\right]+T_{\alpha}'\left[\cg_1\;,\;\check{\vec{\sigma}}_{||}\cg_1\check{\vec{\sigma}}_{||}\right]  \nonumber \\
&+i\sqrt{T_{\alpha}''T}\left[\cg_1\,,\,\check{\vec{\sigma}}_{||}\left\{\cg_1,\cg_2\right\}\check{\vec{\sigma}}_{||}\right].
\label{eq:bcred}
\end{align}
For simplicity we will in the following set $T_{\alpha} = T_{\alpha}'$.

\subsection{SF bilayer}
\begin{figure}[h]
\includegraphics[width=0.5\columnwidth]{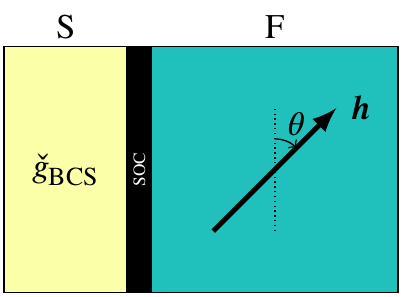}
\caption{The investigated bilayer, consisting of a ferromagnet and a superconductor. There is assumed to be significant spin--orbit coupling at the interface between the two materials, as shown in black. The ferromagnet is modelled by an exchange field $|\vec{h}| = 2\Delta$, pointing in a direction $\theta$ relative to the interface.}
\label{fig:SF}
\end{figure}

As a first example, we consider a bilayer consisting of a superconductor and a ferromagnet, as shown in \cref{fig:SF}. The exchange field in the ferromagnet is directed at an angle $\theta$ from the plane of the interface, with a strength of $|\vec{h}| = 2\Delta$. In this case the Usadel equation takes the following form,

\begin{align}
D\nabla\cdot\cg\nabla\cg + i\left[\varepsilon\rho_3 - \vec{h}\cdot\check{\vec{\sigma}}\;,\;\cg\right] = 0.
\label{eq:husadel}
\end{align}
When both spin--orbit coupling and a magnetization is present in the system, this may lead to long range spin triplet superconducting correlations. Indeed, such correlations were found in a previous work which considered a similar system~\cite{jacobsen_prb_2015}. There, a ferromagnet with Rashba spin--orbit coupling was attached to a superconductor via conventional spin independent boundary conditions. Long range spin triplet correlations were then observed as a zero energy peak in the density of states, the size of which depended upon the angle of the exchange field, $\theta$. Here, we seek to explore whether similar results emerge when the sole contribution to the spin--orbit coupling stems from the boundary conditions.
To quantify the presence of long range spin triplet correlations, we compute the density of states, which is given as
\begin{align}
\nu(\Rv,\varepsilon) = \frac{1}{2}N_0\Re \left(g_{\uparrow\uparrow}(\Rv,\varepsilon) + g_{\downarrow\downarrow}(\Rv,\varepsilon) \right),
\label{eq:dos}
\end{align}
where $N_0$ is the density of states at the Fermi level, and $g_{\sigma\sigma}(\Rv,\varepsilon)$ are spin components of the normal Green function. The presence of long range spin triplets can be inferred from density of states at $\varepsilon = 0$, at which point \cref{eq:dos} may be expressed in terms of the contributions from the anomalous Green function $f = \left(f_s I + \vec{f}_t\cdot\vec{\sigma}\right)i\sigma_y$. With this particular parametrization, where the scalars $f_s$ and $\vec{f}_t$ give the singlet and triplet parts of $f$, respectively, \cref{eq:dos} takes the form
\begin{align}
\nu(\Rv, 0) = 1 - \frac{1}{2}\left|f_s\right|^2 + \frac{1}{2}\left|\vec{f}_{||}\right|^2 + \frac{1}{2}\left|\vec{f}_{\perp}\right|^2.
\label{eq:dos0}
\end{align}
For an exchange field direction indicated by the unit vector $\hat{h}$, the triplet correlation is decomposed into a parallel, $\vec{f}_{||} = \left(\vec{f}_t\cdot\hat{h}\right)\hat{h}$, and an orthogonal, $\vec{f}_{\perp} = \vec{f}_t - \vec{f}_{||}$, component. The motivation for this decomposition is that the spin expectation value of the triplet Cooper pairs is given as $\left\langle \vec{S}\right\rangle \propto i \vec{f}_t(\varepsilon)\times\vec{f}_t^*(-\varepsilon)$~\cite{leggett_rmp_1975,linder_np_2015}. The triplet Cooper pairs whose spins are aligned with the exchange field, will not experience a pair breaking effect, in contrast to orthogonal spin alignments, and hence $\vec{f}_{||}$ and $\vec{f}_{\perp}$ expresses the short and long range triplet correlations, respectively. From \cref{eq:dos0} it is clear that while the presence of singlet superconducting correlations causes a suppression of the density of states, triplet correlations lead to an increase, and thus a potential for the formation of a zero energy peak.
\begin{figure}[h]
\includegraphics[width=\columnwidth]{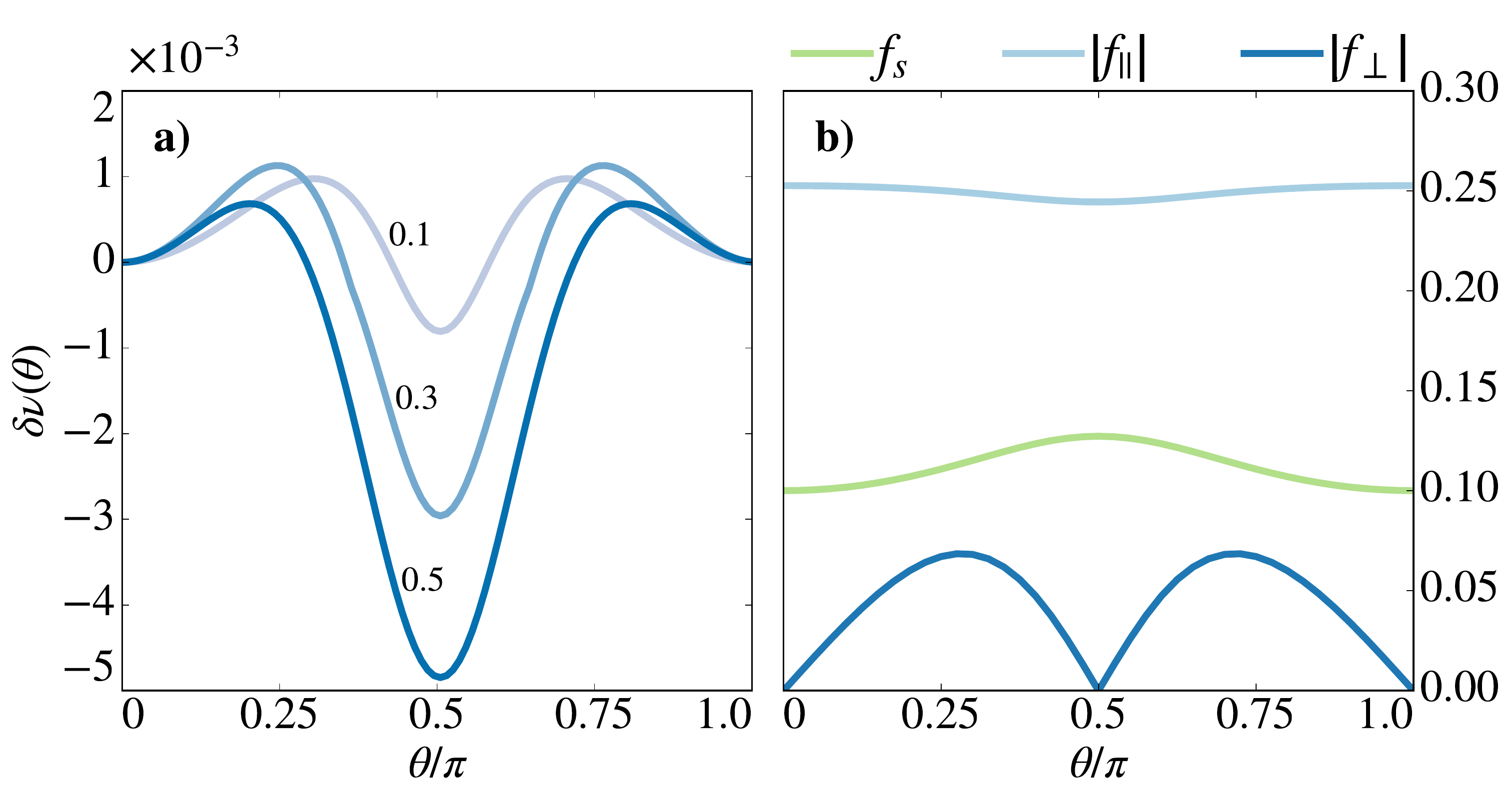}
\caption{Evidence of long range triplet superconducting correlations. a) The maximum density of states in the ferromagnet at zero energy, $\delta\nu(\theta) \equiv \nu(\varepsilon=0,\Rv_{\text{max}},\theta)-\nu(\varepsilon=0,\Rv_{\text{max}},0)$, as a function of the angle of the exchange field $\theta$ relative to the interface. The strength of the spin--orbit coupling at the interface, $T_{\alpha}$, is indicated on the figure. b) The anomalous Green function components as a function of $\theta$ at $T_{\alpha}=0.5$. $f_s$ is the singlet contribution, and $\vec{f}_{||}$ and $\vec{f}_{\perp}$ are the parallel and orthogonal parts of the triplet contribution, respectively. } 
\label{fig:SF_dos}
\end{figure}

\cref{fig:SF_dos}a) shows the change in the density of states at zero energy as the exchange field is rotated away from the interface, $\delta\nu(\theta) \equiv \nu(\varepsilon=0,\Rv_\text{max},\theta)-\nu(\varepsilon=0,\Rv_{\text{max}},0)$, where $\Rv_{\text{max}}$ is the location at which the maximum density of states is found.  A modulation of the zero energy peak is found, similar to the results of Ref.~\cite{jacobsen_prb_2015}. At $\theta = 0$ and $\theta = \pi$, i.e., for an exchange field parallel to the interface, the $T_{\alpha}'$-dependent terms of \cref{eq:bcred} do not contribute, and the boundary conditions reduces to a conventional, spin independent tunnelling barrier. As $\theta$ is increased from zero, so too is the zero energy peak of the density of states, indicating the generation of triplet Cooper pairs. However, as $\theta$ approaches $\pi/2$, a dip is found instead. These results are further elucidated in \cref{fig:SF_dos}b), which shows the angular dependence of the singlet and triplet correlations. The largest modulation is clearly seen in the long range triplets, $\vec{f}_{\perp}$, which is nonzero only when the exchange field has both an in-plane and an out-of-plane component with respect to the interface. In other words, it vanishes for $\theta\in\left\{0,\frac{\pi}{2},\pi\right\}$, in agreement with Ref.~\cite{jacobsen_prb_2015}. Interestingly, it is not purely sinusoidal, but has maxima that are slightly tilted towards $\theta = \pi/2$. A small angular dependence in the singlet, $f_s$, and the short range triplet, $\vec{f}_{||}$, is also observed. At $\theta = \pi/2$, $f_s$ has a slight increase, whereas $|\vec{f}_{||}|$ decreases. This explains the reduction in the zero energy peak of the density of states at $\theta = \frac{\pi}{2}$.

\subsection{SFS Josephson weak link}
\begin{figure}[h]
\includegraphics[width=0.75\columnwidth]{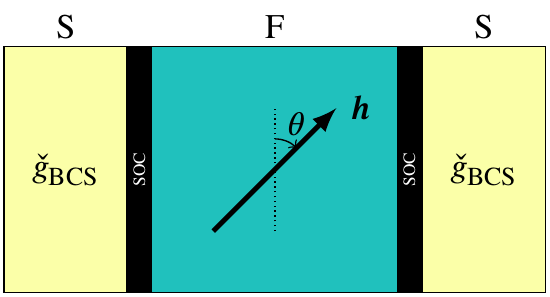}
\caption{A Josephson weak link, where a ferromagnet is sandwiched between two superconductors, with spin--orbit coupling present at the interfaces. The exchange field $\vec{h}$ is directed at an angle $\theta$ relative to the transverse direction of the junction, and has a strength of $|\vec{h}| = 2\Delta$. The distance between the superconductors is assumed to be $L = 2\xi$, where $\xi$ is the superconducting coherence length.}
\label{fig:SFS}
\end{figure}

When two superconductors are separated by a non-superconducting material, they may form a Josephson weak link. When a phase difference $\Delta\phi$ is induced between the superconductors of such systems, for instance by applying a current bias, dissipationless charge currents will flow between them, mediated by the Cooper pairs present in the nonsuperconducting material due to the proximity effect. It is well known that when the intermediary layer consists of a ferromagnet with an inhomogeneous magnetization, long range triplet Cooper pairs may be generated, which are spin polarized, and hence may carry a dissipationless spin current. It was recently predicted that a spin current may also emerge in homogeneous ferromagnets if thin normal metal layers with strong spin--orbit coupling are added between the ferromagnet and the superconductors~\cite{jacobsen_scirep_2016}. To achieve this, the spin--orbit coupling was introduced in thin separate layers, coupled to the surrounding layers by tunnelling barriers, within which the Usadel equation was solved. While experimental verification of these results has proven elusive~\cite{satchell_prb_2018,satchell_arxiv_2019}, the theoretical predictions provide an excellent benchmark for the new boundary conditions, as similar results should be obtained when the spin--orbit coupling is introduced as an interface effect. To verify this, we consider the system illustrated in \cref{fig:SFS}. A homogeneous exchange field is defined in the ferromagnet, with a strength of $|\vec{h}|$, pointing in a direction $\theta$ relative to the transversal direction of the weak link. For this system, we compute the spin supercurrent, which in equilibrium is found from the Green function as
\begin{align}
\vec{I}_s = I_{s,0}\int d\varepsilon\; \Re\text{Tr}\left[\hat{\rho}_3\hat{\vec{\sigma}}\left(\hat{g}\partial_x\hat{g}\right)^K\right]\tanh\frac{\beta\varepsilon}{2},
\end{align} 
where it has been assumed that junction is aligned along the $x$ axis, and $I_{s,0} = \frac{N_0DW}{8}$, for a given junction width $W$, and $\beta = 1/k_{\text{B}}T$, with $T$ the temperature.

\begin{figure}[h]
\includegraphics[width=\columnwidth]{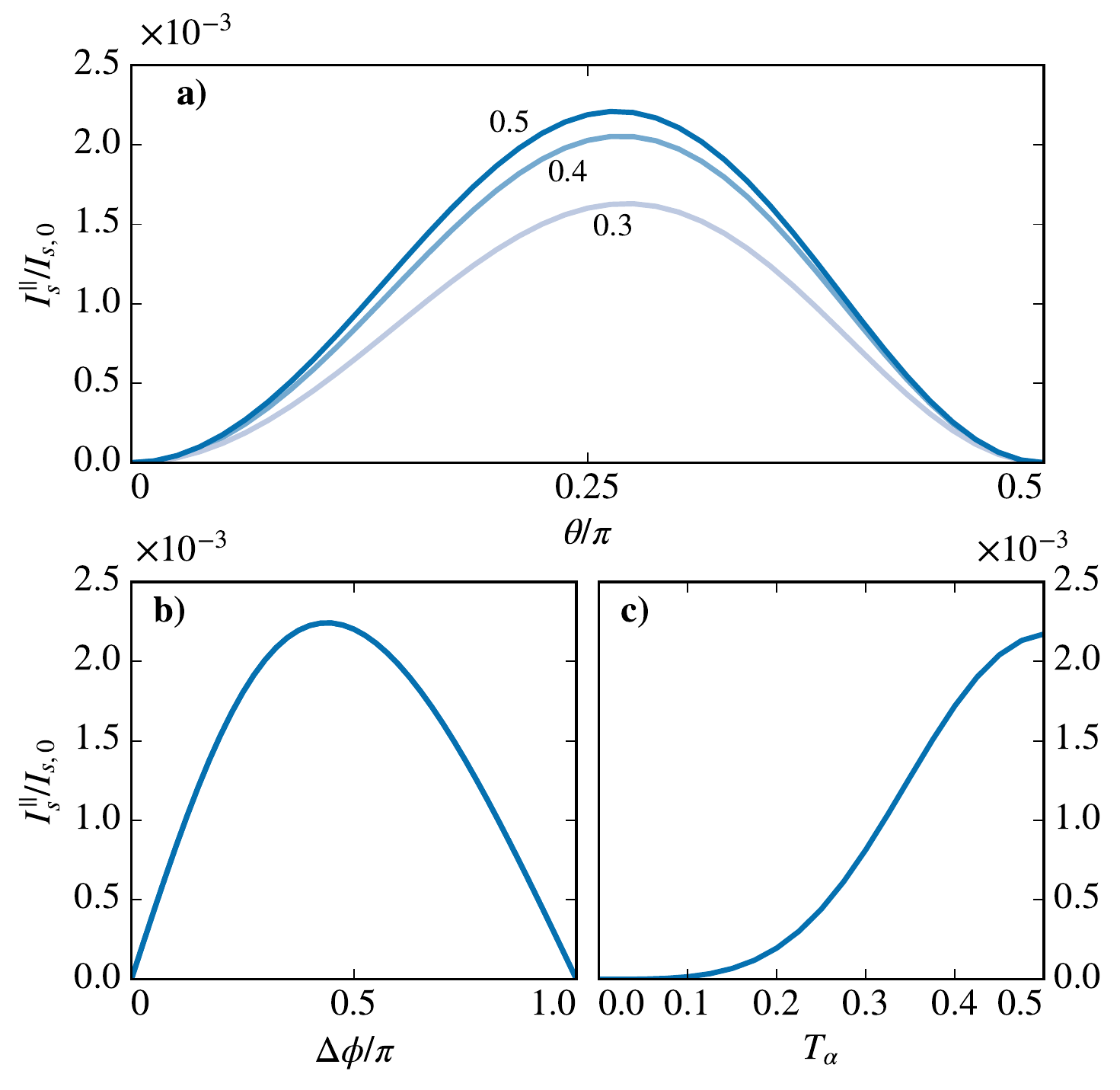}
\caption{The spin current component parallel to the exchange field in the SFS Josephson weak link, scaled by $I_{s,0} = N_0DW/8$. \textbf{a)} shows its variation with the canting angle $\theta$, where the strength of the interface spin--orbit coupling $T_{\alpha}$ is indicated on the figure, \textbf{b)} shows its dependence on the phase difference between the superconductors, $\Delta\phi$, and \textbf{c)} shows the maximum spin current as a function of the interface spin--orbit coupling.}
\label{fig:sfs_dos}
\end{figure}

The results are given in \cref{fig:sfs_dos}, which shows the spin current component aligned parallel to the exchange field, in other words the spin current induced by the long range triplets. In \cref{fig:sfs_dos}a) its dependence on the canting angle $\theta$ is shown. It is noticed that the spin current goes to zero for $\theta = 0$ and for $\theta = \frac{\pi}{2}$. This means that an exchange field with both an in-plane and an out-of-plane component is required in order to observe an effect, similarly to the SF bilayer. In \cref{fig:sfs_dos}b) we show the dependence of the spin current on the phase difference $\Delta\phi$ between the superconductors. It is seen that the current phase relation is approximately sinusoidal, similar to the conventional Josephson effect, indicating that the charge currents have become spin polarized. Finally, we show in \cref{fig:sfs_dos}c) the maximum spin current as a function of the interface spin--orbit coupling $T_{\alpha}$. For low values of the spin--orbit coupling, the spin current has an approximately parabolic form, but reaches a plateau as $T_{\alpha}$ approaches 0.5, the maximum value investigated in this study. A possible interpretation of this is that we are nearing the edge of the domain of validity for the small angle approximation used in the derivation of the boundary conditions, which requires $T_{\alpha}$ to be small. 

\section{Conclusion}
We have derived a new set of boundary conditions for systems in which there is large spin--orbit coupling. This allows the study of, for instance, superconducting hybrid structures with thin heavy metal layers. We demonstrate the use of these boundary conditions by considering an SF bilayer and an SFS Josephson weak link. In both cases we find that whenever the exchange field of the ferromagnet has both an in-plane and an out-of-plane component, long range triplet superconductivity is induced. The findings reported herein are consistent with results found in previous works, where the spin--orbit coupling is approximated by other means.

\begin{acknowledgments}
The authors are grateful to M. Eschrig for valuable input. The authors also wish to thank V. Risingg{\aa}rd,  J. A. Ouassou, and J. R. Eskilt for useful discussions. J. L. acknowledges funding from the Research Council of Norway Center of Excellence Grant Number 262633, Center for Quantum Spintronics, and from the Research Council of Norway Grant No, 240806. J. L. and M. A. also acknowledge funding from the NV-faculty at the Norwegian University of Science and Technology.
\end{acknowledgments}

\end{document}